\pdfoutput=1
\documentclass[10pt,conference]{IEEEtran}

\newif\ifANON
\ANONfalse

\newif\ifDEBUG
\DEBUGfalse

\ifDEBUG
    \newcommand{\WJ}[1]{\textcolor{blue}{[Wenxin: #1]}}
    \newcommand{\MIK}[1]{\textcolor{red}{[Mik: #1]}}
\else
    \newcommand{\WJ}[1]{}
    \newcommand{\MIK}[1]{}
\fi

\usepackage{cite}
\usepackage{amsmath,amssymb,amsfonts}
\usepackage{graphicx}
\usepackage{booktabs}
\usepackage{array}
\usepackage{xcolor}
\usepackage{tikz}
\usetikzlibrary{positioning,arrows.meta,fit,backgrounds}
\usepackage{xurl}
\usepackage[hidelinks]{hyperref}

\newtheorem{definition}{Definition}
\newtheorem{proposition}{Proposition}
\usepackage[skins]{tcolorbox}
\usepackage{fontawesome5}
\newcommand{\icron}{{\color{cbgslate}\faIcon[regular]{clock}}}
\newcommand{\iboard}{{\color{black!70}\faIcon{clipboard-list}}}
\newcommand{\igov}{{\color{orange!70!black}\faIcon{shield-alt}}}
\newcommand{\icli}{{\color{cbgslate}\faIcon{terminal}}}
\newcommand{\iagent}{{\color{cbgslate}\faIcon{robot}}}
\newcommand{\obs}[1]{\textbf{\color{cbgslate}O#1}}
\newcommand{\hyp}[1]{\textbf{\color{cbgteal}H#1}}
\definecolor{cbgslate}{RGB}{70,96,132}
\definecolor{cbgteal}{RGB}{38,112,108}
\tcbset{cbg/.style 2 args={enhanced,frame hidden,sharp corners,boxrule=0pt,boxsep=0pt,
  grow to left by=1.5pt,borderline west={1.5pt}{0pt}{#1},colback=#2,
  left=4.5pt,right=3pt,top=3pt,bottom=3pt,
  before skip=3pt plus 4pt minus 1pt,after skip=3pt plus 4pt minus 1pt,
  fontupper=\small,before upper={\lineskiplimit=1pt\lineskip=1pt}}}
\newtcolorbox{cbgkeybox}{cbg={cbgslate}{blue!4}}
\newcommand{\cbglabel}[3]{\textbf{\color{#1}#2\if\relax\detokenize{#3}\relax\else\ (#3)\fi.}\ \ignorespaces}

\begin{document}
\bstctlcite{IEEEexample:BSTcontrol}

\title{Instant Software Factories}

\ifANON
  \author{\IEEEauthorblockN{Anonymous Author(s)}}
\else
  
\author{
\IEEEauthorblockN{Wenxin Jiang, Mikola Lysenko}
\IEEEauthorblockA{
\textit{Socket Inc.}\\
\{wenxin,mik\}@socket.dev
}
}

\fi

\maketitle

\clubpenalty=10000
\widowpenalty=10000
\displaywidowpenalty=10000
\begin{abstract}

Coding agents can quickly build prototypes, but a long tail of fixes, optimizations, tests, and releases separates a prototype from a product. A \emph{software factory} handles this ongoing work through a repeatable process. We ask whether this work needs dedicated orchestration and how such factories should be judged. Given a tested repository and a coding agent, an \emph{instant software factory} uses only two coordination inputs, a swarm of \emph{cron-job agents} with standing objectives and a shared \emph{communication board}, such as GitHub, where agents coordinate through the traces they leave.
We developed a bootstrap prompt designed to set up both inputs from one round of human input, with governance as a separable layer. We distilled it from two manually configured industrial factories, observed for 15 and 17 days, and have not yet evaluated it. In the GitHub-coordinated CLI factory, daily merges rose 15$\times$ during agent rollout, an observed association. Its agents also filed 95\% of new issues and authored 40 of 58 PRs that changed CI workflows, and at least 5.6\% of issues marked as fixed were later found to have incorrect fixes. Unlike an agent scored on a fixed benchmark, a factory thus changes its own workload, evaluator, and measured progress.
We formulate factory development as multi-objective optimization under this \emph{endogenous feedback}. Over five months, we plan to evolve both factories, test the bootstrap on new repositories, and test five falsifiable hypotheses.

\end{abstract}

\section{Introduction}

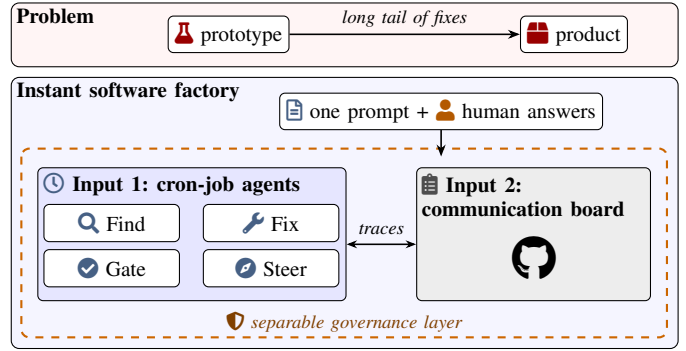
\begin{figure}[t]
\centering
\begin{tikzpicture}[
  font=\footnotesize,
  box/.style={draw, rounded corners=2pt, align=center, inner sep=2.5pt, fill=white},
  mini/.style={box, minimum height=14pt, text width=46pt},
  arr/.style={-{Stealth[length=1.6mm]}, semithick},
  both/.style={{Stealth[length=1.4mm]}-{Stealth[length=1.4mm]}, semithick},
  band/.style={draw, rounded corners=3pt, minimum width=251pt, anchor=north west},
  ttl/.style={anchor=north west, inner sep=2pt, font=\footnotesize\bfseries}
]
\node[band, minimum height=24pt, fill=red!4] (P) at (0,0) {};
\node[ttl] at (P.north west) {Problem};
\node[box] (proto) at (82pt,-12pt) {{\color{red!60!black}\faIcon{flask}}\ prototype};
\node[box] (prod) at (212pt,-12pt) {{\color{red!60!black}\faIcon{box}}\ product};
\draw[arr] (proto) -- node[above=-1pt, font=\scriptsize\itshape]{long tail of fixes} (prod);
\node[band, minimum height=102pt, fill=blue!4] (Q) at (0,-28pt) {};
\node[ttl] at (Q.north west) {Instant software factory};
\node[box] (boot) at (162pt,-41pt) {{\color{cbgslate}\faIcon[regular]{file-alt}}\ one prompt + {\color{orange!70!black}\faIcon{user}}\ human answers};
\draw[orange!80!black, dashed, thick, rounded corners=3pt] (4pt,-55pt) rectangle (247pt,-126pt);
\node[anchor=south, inner sep=2pt, font=\scriptsize\itshape, text=orange!50!black] at (125.5pt,-126pt) {\faIcon{shield-alt}\ separable governance layer};
\node[draw, rounded corners=2pt, fill=blue!10, anchor=north west, minimum width=116pt, minimum height=50pt] (I1) at (10pt,-62pt) {};
\node[anchor=north west, inner sep=2pt, font=\footnotesize\bfseries] at (I1.north west) {{\color{cbgslate}\faIcon[regular]{clock}}\ Input 1: cron-job agents};
\node[mini] (find) at (38pt,-83pt) {{\color{cbgslate}\faIcon{search}}\ Find};
\node[mini] (fix) at (98pt,-83pt) {{\color{cbgslate}\faIcon{wrench}}\ Fix};
\node[mini] (gate) at (38pt,-100pt) {{\color{cbgslate}\faIcon{check-circle}}\ Gate};
\node[mini] (steer) at (98pt,-100pt) {{\color{cbgslate}\faIcon{compass}}\ Steer};
\node[draw, rounded corners=2pt, fill=black!7, anchor=north east, minimum width=88pt, minimum height=50pt] (I2) at (241pt,-62pt) {};
\node[anchor=north west, inner sep=2pt, font=\footnotesize\bfseries, align=left] at (I2.north west) {{\color{black!70}\faIcon{clipboard-list}}\ Input 2:\\communication board};
\node at (197pt,-96pt) {\LARGE\faIcon{github}};
\draw[both] (I1.east |- 0,-91pt) -- node[above, font=\scriptsize\itshape]{traces} (I2.west |- 0,-91pt);
\draw[arr] (162pt,-48pt) -- (162pt,-58pt);
\end{tikzpicture}
\caption{Instant software factories. Cron-job agents and a communication board such as GitHub coordinate the factory. Governance is separable but mandatory.}
\label{fig:overview}
\end{figure}

Coding agents, systems based on large language models (LLMs) that read code, use tools, and propose changes, have opened over 930,000 pull requests (PRs) on GitHub~\cite{li2026aidev} and quickly build prototypes. Yet a prototype is far from a product. Bugs, edge cases, regressions, compatibility breaks, flaky continuous integration (CI), and releases create a long tail of work (Fig.~\ref{fig:overview}, top). Prompting an agent through every task makes developer attention the bottleneck. A \emph{software factory} handles such work with specialized roles, common tools, and repeatable processes~\cite{cusumano1989factory}.

Building a factory for this tail is often treated as a substantial engineering effort. Stripe, Ramp, Spotify, and Uber have built internal agent platforms~\cite{gray2026minions2,ramp2026inspect,charas2025honk1,uber2026factory}, research systems use orchestrators and message passing~\cite{qian2024chatdev,hong2024metagpt,lyu2026thebotcompany}, and MAGE models factories as production systems with many controls~\cite{davis2026magebook}. 
But how much infrastructure does a software factory actually need, and how should its progress be judged?

We explore a simpler alternative, an \emph{instant software factory}, and separate two claims. Our \emph{sufficiency hypothesis} is that \textbf{two coordination inputs suffice for a factory for PR-sized work}. Our \emph{feasibility hypothesis} is that a bootstrap prompt we developed can configure both inputs in one session from a single round of human answers (Fig.~\ref{fig:overview}, bottom). One factory gives early evidence for the first, and the second is untested.

\begin{figure*}[t]
\centering
\begin{tikzpicture}[
  font=\footnotesize,
  base/.style={rounded corners=2pt, inner sep=0pt, minimum height=12.5pt, anchor=north west},
  box/.style={base, draw, fill=white},
  gbox/.style={base, draw=orange!80!black, dashed, thick, fill=orange!5},
  hbox/.style={base, draw, fill=yellow!18},
  hgbox/.style={base, draw=orange!80!black, dashed, thick, fill=yellow!18},
  arr/.style={-{Stealth[length=1.4mm]}, semithick},
  lbl/.style={font=\scriptsize\itshape, inner sep=1.5pt},
  panel/.style={draw, rounded corners=3pt, anchor=north west, minimum height=118pt, minimum width=166pt},
  ttl/.style={anchor=north west, inner sep=3pt, font=\footnotesize\bfseries},
  tag/.style={circle, fill=cbgslate, text=white, inner sep=0pt, minimum size=8pt, font=\tiny\bfseries},
  gtag/.style={tag, fill=orange!80!black}
]
\newcommand{\ic}[2][cbgslate]{\makebox[14pt][c]{\color{#1}#2}\hspace{1pt}}

\node[panel, fill=blue!4] (A) at (0,0) {};
\node[ttl] at (A.north west) {(a) The prompt's seven parts};
\foreach \i/\st/\tg/\t in {
  1/box/tag/{\ic{\faIcon{users}}\textbf{Roles}: eight standing objectives},
  2/box/tag/{\ic{\faIcon[regular]{clock}}\textbf{Routines}: the manager owns them},
  3/box/tag/{\ic{\faIcon{search}}\textbf{Inspect} the repo, then ask once},
  4/gbox/gtag/{\ic[orange!70!black]{\faIcon{id-badge}}\textbf{Identity} and human approval},
  5/gbox/gtag/{\ic[orange!70!black]{\faIcon{lock}}\textbf{Controls} and a stop switch},
  6/gbox/gtag/{\ic[orange!70!black]{\faIcon{vial}}\textbf{Prove} controls before enabling},
  7/box/tag/{\ic{\faIcon[regular]{sticky-note}}\textbf{Platform notes} from setups}} {
  \node[\st, minimum width=139pt, align=left, text width=135pt] (s\i) at (21pt,{-17pt-(\i-1)*14.5pt}) {\hspace{2pt}\t};
  \node[\tg] at ([xshift=-8pt]s\i.west) {\i};
}

\node[panel, fill=green!5] (B) at (172pt,0) {};
\node[ttl] at (B.north west) {(b) One setup session};
\foreach \j/\st/\t in {
  1/box/{\ic{\faIcon{search}}\textbf{Inspect} the repo, CI, and access},
  2/hbox/{\ic[orange!70!black]{\faIcon[regular]{comments}}\textbf{Ask} the human once},
  3/box/{\ic{\faIcon{cogs}}\textbf{Provision} routines and a setup PR},
  4/gbox/{\ic[orange!70!black]{\faIcon{vial}}\textbf{Prove} the controls},
  5/hgbox/{\ic[orange!70!black]{\faIcon{user-check}}\textbf{Approve} the setup PR},
  6/box/{\ic{\faIcon{play}}\textbf{Enable} the schedules}} {
  \node[\st, minimum width=140pt, align=left, text width=136pt] (b\j) at (185pt,{-17pt-(\j-1)*17.4pt}) {\hspace{2pt}\t};
}
\foreach \a/\b in {b1/b2,b2/b3,b3/b4,b4/b5,b5/b6} {\draw[arr] (\a.south) -- (\b.north);}

\node[panel, minimum width=172pt, fill=orange!5] (C) at (344pt,0) {};
\node[ttl] at (C.north west) {(c) The factory at run time};
\tikzset{role/.style={box, minimum width=70pt, align=center}}
\node[role] (find)  at (352pt,-16pt) {\ic{\faIcon{search}}\textbf{Find}};
\node[role] (fix)   at (438pt,-16pt) {\ic{\faIcon{wrench}}\textbf{Fix}};
\node[box, minimum width=156pt, align=center, fill=black!7] (board) at (352pt,-41pt) {\ic[black]{\faIcon{github}}\textbf{Communication board}};
\node[role] (steer) at (352pt,-66pt) {\ic{\faIcon{compass}}\textbf{Steer}};
\node[role] (gate)  at (438pt,-66pt) {\ic{\faIcon{check-circle}}\textbf{Gate}};
\draw[arr] (find.south) -- node[lbl, right]{issue} (find.south |- board.north);
\draw[arr] ([xshift=-9pt]fix.south |- board.north) -- node[lbl, left]{issue} ([xshift=-9pt]fix.south);
\draw[arr] ([xshift=9pt]fix.south) -- node[lbl, right]{PR} ([xshift=9pt]fix.south |- board.north);
\draw[{Stealth[length=1.4mm]}-{Stealth[length=1.4mm]}, semithick] (steer.north) -- node[lbl, right]{assign} (steer.north |- board.south);
\draw[arr] ([xshift=-9pt]gate.north |- board.south) -- node[lbl, left]{PR} ([xshift=-9pt]gate.north);
\draw[arr] ([xshift=9pt]gate.north) -- node[lbl, right]{merge} ([xshift=9pt]gate.north |- board.south);
\node[box, minimum width=156pt, align=center, fill=black!2] (loop) at (352pt,-86pt) {\ic{\faIcon[regular]{clock}}\textbf{Each run}: wake, read, act once, exit};
\node[gbox, minimum width=156pt, align=center] (govc) at (352pt,-101pt) {\ic[orange!70!black]{\faIcon{shield-alt}}{\color{orange!50!black}\textbf{Governance}: approval, stop switch}};
\end{tikzpicture}
\caption{Anatomy of the bootstrap. (a) Prompt parts. (b) Setup session. (c) Running factory. Orange dashed items form the separable governance layer.}
\label{fig:bootstrap}
\end{figure*}
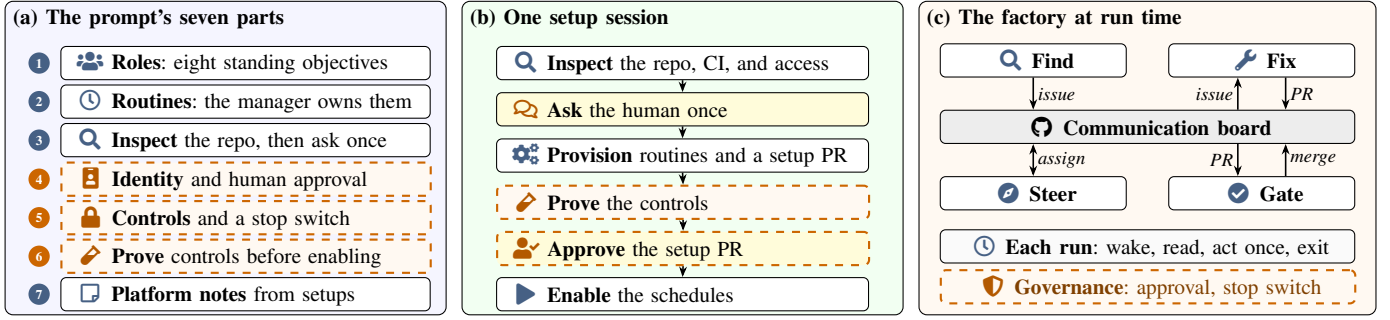

We distilled this design from two industrial factories built by hand before the bootstrap existed. They suggest that judging a factory is harder than assembling one, which we see as the main new problem. Agent benchmarks fix the tasks and tests, and studies of reward overoptimization fix the reward model~\cite{gao2022overoptimization}. A factory's actions instead change its product, the work it is given, and the process that evaluates it, a loop we call \emph{endogenous feedback}. In the CLI factory, agents filed 95\% of issues (\emph{endogenous workload}) and authored 40 of 58 CI workflow PRs (\emph{endogenous evaluation}), and 7 incorrect fixes shipped unrepaired (\emph{delayed verdicts}). Throughput can then overstate progress, so factories must optimize independently valued, validated, and delivered changes.

\begin{cbgkeybox}
\textbf{\color{cbgslate}Key idea.}
Given a tested repository with CI and a coding agent with repository access, a factory for PR-sized work may need only two coordination inputs:
(1)~\icron\ \textbf{Cron-job agents} with standing objectives, such as finding bugs or improving CI. Each run reads the board, does one bounded task, and exits.
(2)~\iboard\ \textbf{A communication board}, such as GitHub's issues, PRs, checks, and reviews. Agents coordinate through \emph{stigmergy}, following each other's traces without an orchestrator.
\igov\ \textbf{Governance}, such as approval and a stop switch, is a separable, mandatory layer scaled to risk.
\end{cbgkeybox}

We contribute (1)~a formulation of factory development as \textbf{multi-objective optimization under endogenous feedback}, with a progress measure that agents cannot edit and five testable hypotheses (\S\ref{sec:discussion}--\ref{sec:future}), (2)~\textbf{observations} of endogenous workload, endogenous evaluation, and delayed verdicts in two factories (\S\ref{sec:evidence}), and (3)~a \textbf{two-input design}, posed as a sufficiency hypothesis, with a released, not yet evaluated \textbf{bootstrap prompt}~\cite{lysenko2026instantfactory} (\S\ref{sec:bootstrap}).

\section{Agent Swarms and Software Factories}
\label{sec:related}

\noindent\textbf{Swarms and stigmergy.}
Social insects coordinate through \emph{stigmergy}, leaving traces in a shared environment that guide later actions~\cite{bonabeau1999swarm}, and blackboard systems let specialists coordinate through shared information~\cite{erman1980hearsay}. LLM agents sharing persistent artifacts develop specialized roles~\cite{pal2026swarmworld}, but teams without centralized verification can propagate errors~\cite{kim2025scaling}, motivating a shared board and a merge gate.

\noindent\textbf{Agent factories.}
Research systems and toolkits rely on message passing or dedicated orchestrators~\cite{qian2024chatdev,hong2024metagpt,yegge2026gastown,lyu2026thebotcompany}. Practitioners have also coordinated scheduled agents through GitHub issues and PRs~\cite{githubnext2025aw,syme2026peli,bichard2026ona}, but typically configure agents individually and report activity or merge rates. We add no coordination mechanism and instead test whether these primitives suffice for a whole factory initialized by one prompt, judging factories under endogenous feedback by changes that hold after merge and reach users.

\section{The Bootstrap}
\label{sec:bootstrap}

\begin{figure*}[t]
\centering
\definecolor{fblue}{HTML}{2A78D6}\definecolor{forange}{HTML}{EB6834}\definecolor{fgrey}{HTML}{6F6E69}
\includegraphics[width=\textwidth]{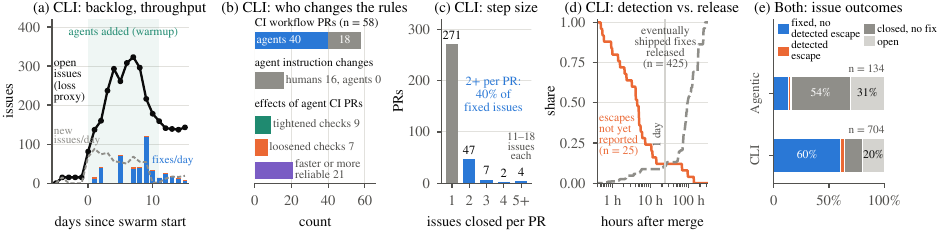}
\caption{The two factories through an optimization lens. (a) CLI open issues (loss proxy), \textcolor{fgrey}{new issues per day}, and daily fixed issues (throughput) with \textcolor{fblue}{no detected escape} or an \textcolor{forange}{escape}. (b) CLI changes to CI workflows and agent instructions by \textcolor{fblue}{agents} and \textcolor{fgrey}{humans}, and LLM-rated effects of agent CI PRs. (c) CLI step size. (d) CLI detection against release, as the share of 25 \textcolor{forange}{escapes not yet reported} and, for a different cohort, of 425 \textcolor{fgrey}{eventually shipped fixed issues (escapes included) already released}. (e) Outcomes of 704 CLI issues and 134 agentic defect types. No detected escape does not establish correctness.}
\label{fig:evidence}
\end{figure*}

We developed a bootstrap prompt, distilled from two hand-built factories (\S\ref{sec:evidence}), to set up a factory in one session, an untested feasibility hypothesis (\S\ref{sec:future}). A developer gives a coding agent in the target repository a 1,400-word prompt (Fig.~\ref{fig:bootstrap}). The design targets tasks that fit one PR and can wait an hour. Production deployment and multi-repository work are out of scope, and tightly coupled work may need planners~\cite{lin2026scaling}.

\noindent\textbf{\icron\ Input 1: cron-job agents.}
Each agent is a saved \emph{routine} launched on a cron schedule at most once per hour~\cite{anthropic2026routines}.
 Runs keep no memory beyond the board. Eight roles divide the work. QA and performance agents \emph{find} problems and file issues. Implementation, CI, and architecture agents \emph{fix} them through PRs. Review and merge agents \emph{gate} changes. A manager, itself a routine, \emph{steers} work through the board, assigning tasks, tuning models and schedules within budget, and balancing objectives (\S\ref{sec:discussion}).

\noindent\textbf{\iboard\ Input 2: the board.}
Issues record tasks and evidence, PRs carry changes, and checks and reviews provide verdicts. Any board that holds these three kinds of traces suffices. Our bootstrap targets GitHub and scheduled Claude Code tasks.

\noindent\textbf{{\color{cbgslate}\faIcon[regular]{file-alt}} One setup session.}
The agent inspects the repository, asks the human once about budget, QA priorities, and approvers, and opens a setup PR, enabling schedules only after testing governance controls and obtaining approval.

\noindent\textbf{Manual setup lessons.}
Hand-built routines shaped the platform notes. Tool lists alone did not enforce GitHub permissions, so the prompt tests each restriction. CI capacity and GitHub reliability were the main bottleneck, motivating adaptive run rates (\S\ref{sec:discussion}).

\noindent\textbf{\igov\ A separable governance layer.}
Governance restricts agent actions without changing how they coordinate. For software that users install, merges by default require passing checks, independent review of the latest commit, and human approval. Agents check a stop switch before every write and must ignore instructions in retrieved content. Teams scale controls to the risks of their changes, but the layer remains mandatory.

\section{Two Working Factories}
\label{sec:evidence}

We examine four dimensions of factory operation, \emph{continuity} (working without repeated prompting), \emph{governance} (controlling agent actions by risk), \emph{correctness} (changes that hold under independent checks), and \emph{delivery} (changes reaching users).
Following case-study and GitHub-mining guidelines~\cite{runeson2009case,kalliamvakou2014promises}, we studied two industrial factories operated by one organization: a \emph{CLI system} for 15 days (escapes tracked through day~17) and an \emph{agentic system} for 17 days. The CLI factory instantiates the two-input design, and the agentic factory offers a contrasting architecture. Neither evaluates the bootstrap's automated setup.

\noindent\textbf{\icli\ The CLI system} applies security patches to dependencies across nine software package registries and about twenty package managers. It uses exactly two coordination inputs, about 30 scheduled agents, introduced over days 0--11, and GitHub as their shared board, with no orchestrator.

\noindent\textbf{\iagent\ The agentic system} patches vulnerabilities in open-source packages, validating each patch with a proof-of-concept attack and regression tests. We study the swarm maintaining its validation \emph{harness}, whose checks returned no verdict in 28\% of reruns. It instead uses service loops and a shared database work list. A scout groups failures into defect types, a fixer submits PRs, a verifier reruns checks, a judge rejects package-specific fixes, and a recovery agent restarts blocked runs.

\noindent\textbf{{\color{cbgslate}O1 (continuity):} both factories sustained work with different architectures.}
After the CLI swarm started, daily merged PRs rose from 2.2 to about 35, peaking at 115 on day~9. Agents filed 95\% of 704 new issues, and merged PRs closed 450 (64\%) as \emph{fixed}~\cite{guo2010fixed} within a median of 46 hours (Fig.~\ref{fig:evidence}e). Open issues, a proxy for loss, peaked at 323 as agents discovered bugs, then fell to about 140 and stabilized as new reports matched fixes (Fig.~\ref{fig:evidence}a). On a sampled day with 69 merges, agents cost \$1.2k, or \$17 per merge before CI costs. The agentic factory spent \$869 on 21 merged harness fixes, or \$41 each, resolving 21 of 134 defect types (16\%).

\noindent\textbf{{\color{cbgslate}O2 (governance):} controls depend on context and may limit throughput.}
In the CLI factory, 77\% of agent merges changed user-installed code. A second person approved 534 of 541 merged PRs. A merge queue added on day~7 rejected 29 failing agent PRs and delayed others a median 26 minutes as daily merges doubled. Agentic fixes instead affect only an internal harness, and a judge screens them before human approval. One reviewer approved 502 CLI PRs without requesting changes, questioning review effectiveness and, as another agent-first team found~\cite{lopopolo2026harness}, scalability.

\noindent\textbf{{\color{cbgslate}O3 (governance):} agents also modify the rules.}
CLI agents authored 40 of 58 PRs changing CI workflows, while the maintainer made all 16 changes to agent instructions. Two LLM raters classified CI effects ($\kappa=0.91$), including 7 PRs that loosened checks (Fig.~\ref{fig:evidence}b). Humans reviewed these changes, so edits to the factory's own controls stayed governed.

\noindent\textbf{{\color{cbgslate}O4 (correctness):} merged fixes do not always hold.}
By day~17, at least 25 of 450 fixed CLI issues (5.6\%) were \emph{escapes}: incorrect fixes~\cite{yin2011fixes} that broke another case of the issue (17) or left part unfixed (8)~\cite{park2012supplementary}. None was reopened, although reopening often signals faulty fixes~\cite{shihab2013reopened}. In the agentic factory, 3 of 14 replayed fixes (21\%) recurred.

\noindent\textbf{{\color{cbgslate}O5 (delivery):} merged fixes must reach users.}
Two CLI releases shipped 94\% of fixed issues after a median \emph{integration delay}~\cite{dacosta2018rapid} of 3.9 days. Agentic harness fixes propagate only when the runs they unblock restart, and a recovery sweep resumed 11 of 22 blocked runs.

\noindent\textbf{{\color{cbgslate}O6 (delivery):} escapes race releases.}
Reports, mostly from bug-hunting agents, identified 22 of 25 CLI escapes within a day of merge (Fig.~\ref{fig:evidence}d). Release delays allowed 14 to be corrected before shipment, but 7 shipped unfixed, including 4 before being reported.

\noindent\textbf{Threats to validity.}
Both factories belong to one organization, the bootstrap's sole source, and were observed for two to three weeks. Both retained human approval, leaving the minimum necessary governance unknown. CLI agents were introduced over days 0--11, and their instructions changed 16 times, so before-and-after differences cannot establish causality. The different workloads and lack of an architectural baseline prevent claims of superiority. Operators stopped the agentic swarm on day~16, and humans amended 7 of its 21 fixes. Escape counts depend on scouting coverage, classifications rely on LLM raters, and CLI costs come from one sampled day. We found no evidence of intentional gaming and did not test whether agent CI changes caused escapes.

\section{The Software Factory as a Multi-Objective Optimization Problem}
\label{sec:discussion}

Inspired by search-based software engineering~\cite{harman2012sbse}, we frame factory configuration as a \emph{multi-objective optimization problem}, with merged PRs as steps and manager-controlled settings (objectives, schedules, PR scope, and notes) as hyperparameters under governance constraints (Table~\ref{tab:mapping}). The framing defines what to measure and test and proposes no new optimizer. Let the factory state be $S_t=(P_t,B_t,E_t)$, the product, backlog, and validation process, with measured progress $\hat{R}_t=E_t(P_t,B_t)$. Agent actions $a_t$ reach $\hat{R}_t$ through three channels. They file issues into $B_t$ (workload, \obs{1}), edit CI so that $E_{t+1}=G(E_t,a_t)$ (evaluator, \obs{3}), and merge before an independent evaluator $E^{*}$ judges $P_t$ (delay, \obs{4}, \obs{6}). Maximizing $\hat{R}_t$ can thus reward growing the backlog, loosening checks, or merging early. We therefore measure progress with $E^{*}$, which agents cannot edit, and counter each channel with a safeguard, preset task values, held-out tests outside $G$ (\hyp{5}), and estimates taken after a fixed delay.

\begin{table}[t]
\centering
\caption{Optimization concepts mapped to factories, with preliminary data from \S\ref{sec:evidence}.}
\label{tab:mapping}
\scriptsize
\setlength{\tabcolsep}{2.0pt}
\renewcommand{\arraystretch}{1.08}
\begin{tabular}{@{}ll@{\hspace{5pt}}ll@{}}
\toprule
\multicolumn{2}{@{}c}{\emph{Concept}} & \multicolumn{2}{c@{}}{\emph{Preliminary data}} \\
\cmidrule(r{4pt}){1-2}\cmidrule(l{2pt}){3-4}
\textbf{Optimizer} & \textbf{Factory} & \textbf{CLI system} & \textbf{Agentic system} \\
\midrule
\textbf{Initialization} & bootstrap & manual, 11-day rollout & manual \\
\textbf{Objectives} & agent goals (\hyp{1}) & $\sim$30 agents & 5 roles \\
\textbf{Learning rate} & run rate (\hyp{2}) & $\sim$30 fixed schedules & service loops \\
\textbf{Step size} & issues per PR (\hyp{3}) & 450/331 = 1.36 & one defect type \\
\textbf{Momentum} & notes (\hyp{4}) & issue and PR threads & shared defect list \\
\textbf{Generalization gap} & escapes (\hyp{5}) & $\ge$25/450 issues & 3/14 recurred \\
\bottomrule
\end{tabular}
\end{table}

\noindent\textbf{Initialization.}
As initialization affects convergence~\cite{sutskever2013momentum}, the bootstrap would initialize objectives, roles, schedules, and controls. Its defaults come from two manual factories, may inherit their biases, and have untested effects.

\noindent\textbf{Objectives (\hyp{1}).}
A single weighted loss can hide trade-offs between competing objectives~\cite{sener2018mtl}. CLI agents accelerated CI by moving end-to-end tests to nightly runs (\obs{3}), delaying detection of breakage until after merge. Specialists may expose such conflicts so the manager can protect the worst-served objective, while a generalist folds them into one loss.

\noindent\textbf{Loss, velocity, and cost.}
Open issues estimate \emph{loss}. We define \emph{velocity} as lasting, delivered progress, like time-to-accuracy in training~\cite{coleman2019dawnbench}, and treat merge rates as throughput only. $N_i(\pi)$ counts unprompted changes toward objective $i$ per week under configuration~$\pi$ that $E^{*}$ judges correct, that reach users, and that carry severity weights set independently of the agents. The manager seeks a Pareto-optimal $\pi$:
\begin{equation}
\max_{\pi}\ \big(N_1(\pi),\dots,N_m(\pi)\big)
\label{eq:l2}
\end{equation}
subject to weekly LLM, CI, and review cost $C(\pi)\le C_{\max}$, human time $T(\pi)\le T_{\max}$, escape rate $\le\tau$, and governance constraints. Hidden escapes make measured $N_i$ an upper bound, taken after a fixed audit window such as two weeks.

\noindent\textbf{Worked example.}
Pooled over the CLI study, throughput counts 450 fixed issues. Delivery keeps the 425 that shipped, and correctness drops the 7 that shipped with a detected escape, leaving at most 418. As agents filed 95\% of issues (\obs{1}), the maintainer would assign each an objective and a severity from a preset rubric, such as broken installs above wrong messages, and divide each $N_i$ by a preset target so objectives are comparable (\hyp{1}). Agent CI PRs would count only toward the CI objective, never as product progress.

\noindent\textbf{Learning rate and momentum (\hyp{2}, \hyp{4}).}
Like an adaptive learning rate~\cite{kingma2015adam}, run frequency could track verified progress, escapes, and cost, which neither factory did, although CI and review limited the pace (\obs{2}). Like momentum~\cite{sutskever2013momentum}, \emph{notes} in issue threads or the defect list carry context across memoryless runs but can mislead when stale.

\noindent\textbf{Step size (\hyp{3}).}
Issues per PR coarsely approximate step size, and each PR incurs review, CI, and merge costs (\obs{2}). Of 331 issue-closing CLI PRs, 60 fixed multiple issues, 40\% of all fixed issues (Fig.~\ref{fig:evidence}c), with a per-issue escape rate of 3.9\% versus 6.6\% for single-issue PRs, an inconclusive gap (Fisher's $p=0.29$, ignoring PR clustering).

\noindent\textbf{Generalization gap (\hyp{5}).}
A patch may pass its own tests but fail held-out ones~\cite{smith2015overfitting}, as some fixes in both factories did (\obs{4}). Bug-hunting agents already act as held-out tests (\obs{6}). \hyp{5} tests whether held-out tests, run before merge outside $G$ and hidden from the agents they judge, reduce escapes more than agent-modified CI checks (\obs{3}).

\section{Future Plans}
\label{sec:future}

\begin{table}[t]
\centering
\caption{Planned experiments at equal cost. H1--H4 test both factories, and H5 only the CLI factory.}
\label{tab:plan}
\scriptsize
\setlength{\tabcolsep}{3pt}
\renewcommand{\arraystretch}{1.12}
\begin{tabular}{@{}l
  >{\raggedright\arraybackslash}p{0.45\columnwidth}
  >{\raggedright\arraybackslash}p{0.38\columnwidth}@{}}
\toprule
\textbf{Hyp.} & \textbf{Comparison (unit)} & \textbf{Prediction} \\
\midrule
\hyp{1} & Specialists vs.\ generalist (area) & Higher min.\ $N_i$/target \\
\hyp{2} & Adaptive vs.\ fixed rates (week) & Higher velocity at equal cost \\
\hyp{3} & Up to 5 vs.\ 1 issue/PR (root cause) & More fixes/review hour, same escapes \\
\hyp{4} & Recent vs.\ all vs.\ no notes read (run) & Recent notes, highest velocity \\
\hyp{5} & Held-out vs.\ standard gate (CLI PR) & Fewer escapes, slowdown $\le\delta$ \\
\bottomrule
\end{tabular}
\end{table}

Software engineering uses case studies to develop and test hypotheses~\cite{runeson2009case}, as Mockus et al.\ did with Apache and Mozilla~\cite{mockus2002twocase}. Over five months, we will evolve both factories and restart the agentic swarm. For \emph{feasibility}, we will run the bootstrap on new repositories, comparing setup effort, control-test failures, and time to the first $E^{*}$-validated merge with manual setup. For \emph{sufficiency}, we will record whether operators must add coordination, such as an orchestrator or a work database, to sustain $N_i$ for a month. \hyp{5}, our central experiment, tests the evaluator channel. \hyp{1}--\hyp{4} score arms by $N_i$, which discounts agent-filed value and unverified merges, so gains cannot come from inflation.

\noindent\textbf{Learn within each factory.}
Each experiment in Table~\ref{tab:plan} randomizes its own unit under existing governance at equal budgets. \hyp{1} alternates each package manager or defect type between specialist and generalist arms across periods, so each area serves as its own control. \hyp{2} randomizes schedule type within consecutive week pairs, modeling trend and carryover. \hyp{3} groups issues by shared root cause before randomization, stratified by group size, with PR-clustered analysis. \hyp{4} hides notes older than a week or all notes from a run without changing its tasks. \hyp{5} randomizes CLI PRs, and arm-blind scouts judge escapes per fixed issue for 14 days after merge, with PR-clustered analysis and a preset velocity margin $\delta$. Detecting a halved 5.6\% escape rate at 80\% power and $\alpha=0.05$ needs 800+ fixes per arm, about three months at late-study rates, more under clustering.

\noindent\textbf{Generalize across factories.}
Settings that hold in both factories and, by architectural similarity~\cite{wieringa2015six}, in factories with public boards~\cite{syme2026peli,bichard2026ona} become candidate best practices.

\noindent\textbf{Domain-specific or general.}
Prior software factories targeted particular product families or domains~\cite{greenfield2003factories,pavlenko2026aifactory}. On DARPA's AIxCC tasks, specialized systems fixed bugs missed by general agents~\cite{zhang2026aixcc}. Since no optimizer performs best on every problem class~\cite{wolpert1997nfl}, we will test whether lessons from our security factories transfer to other domains. Generalizable settings can become bootstrap defaults, and others setup questions.

\noindent\textbf{Self-tuning and evolution.}
Like an experience factory~\cite{basili1992sel}, an \emph{adaptation skill} will let the manager tune settings from cost, correctness, and delivery within governance~\cite{lee2026metaharness,zhang2025dgm,hu2025adas}.

\section*{Data Availability}
The artifact~\cite{lysenko2026instantfactory} contains the bootstrap and will include its revisions and experimental scripts. Case records remain private for confidentiality.

\bibliographystyle{IEEEtran}
\IfFileExists{arxiv/refs.bib}{\bibliography{arxiv/refs}}{\bibliography{refs}}

\end{document}